\documentclass[aps,prb,twocolumn,groupedaddress]{revtex4}

\usepackage{graphicx}
\usepackage{dcolumn}
\usepackage{bm}

\begin{document}


\title{Supercurrent and Conductance Quantization in a Superconducting Quantum Point Contact}


\author{T. Bauch}
 \email{thilo.bauch@mc2.chalmers.se}
\author{E. H\"{u}rfeld}
\author{V.M. Krasnov}
\author{P. Delsing}
\affiliation{Microtechnology and Nanoscience, MC2, Chalmers University of Technology, S-412 96 G\"{o}teborg, Sweden}

\author{H. Takayanagi}
\author{T. Akazaki}
\affiliation{NTT Basic Research Laboratories, 3-1 Morinosato-Wakamiya, Atsugi-Shi, Kanagawa 243-01, Japan}

\date{\today}

\begin{abstract}
We study the quantization of the supercurrent and conductance of a superconducting quantum point contact (SQPC) in a superconductor-two dimensional electrongas-superconductor (S-2DEG-S) Josephson junction with a split gate. The supercurrent and the conductance values change stepwise as a function of the gate voltage. We observe the onset of the first transport mode contributing both to the supercurrent and the conductance of the SQPC. Furthermore the steps in the supercurrent and the conductance appear at the same gate voltage values. This shows, that each transport mode in the SQPC contributes both to the normal state conductance and to the supercurrent.

\end{abstract}

\pacs{}

\maketitle

\section{introduction}
In analogy with the quantized conductance of a normal conducting quantum point contact (QPC) \cite{vWe88,wah88} a quantization of the supercurrent should be observed in a superconducting quantum point contact (SQPC). \cite{bee91,fur91,fur92,cht00,shc00} This results from the quantization of the transverse momentum of the 2DEG quasiparticles in the QPC constriction with a width of the order of the Fermi wave length $\lambda_{F}$. The number of transport modes is given by $2W_{g}/\lambda_{F}$, where $W_{g}$ is the constriction width. Each transport mode contributes one quantized conductance unit $\Delta G_{0}=2e^{2}/h$ to the total conductance and one quantized supercurrent unit $\Delta I_{C0}$ to the critical current of the SQPC. In the limit of a short junction where the length of the junction $L$ is much smaller than the superconducting coherence length $\xi_{0}$ a stepwise change of the supercurrent and conductance was observed in a mechanically controllable break junction.\cite{mul92} The quantization of the critical current $I_{c}$ in the limit of a long junction $L\geq\xi_{0}$ was observed in a ballistic S-2DEG-S Josephson junction where the constriction in the normal part was varied by changing the voltage of the split gate. \cite{tak95b} In Ref. \onlinecite{tak95b} a stepwise change of the supercurrent and the conductance could be observed by varying the gate voltage of the split gate. Surprisingly the position of the critical current step was different from that of the corrensponding conductance step. The present theory predicts an agreement between the position of the conductance and supercurrent steps.
In this paper we present for the first time experimental data of a SQPC in a S-2DEG-S Josephson junction where the onset of the first transport mode contributing both to the supercurrent and the conductance of the SQPC can be observed. In addition the data show a one to one correlation of the supercurrent steps and the conductance steps as a function of gate voltage.


\section{Theory}

\begin{figure}[htb]
	\begin{center}
		\includegraphics[angle=270,width=8cm]{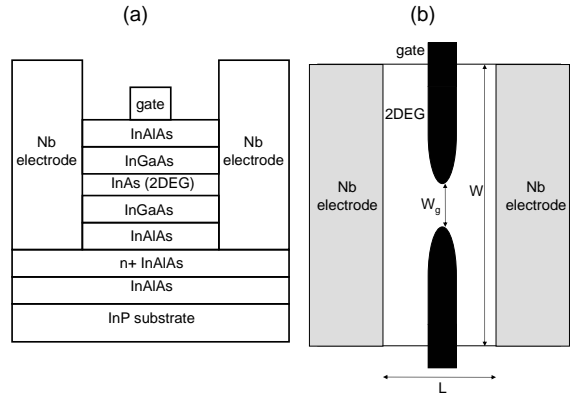}
	\end{center}
	\caption{(a) Cross sectional view of the superconducting quantum point contact (SQPC). The carrier concentration $n_{S}$ in the 2DEG can be varied by applying a voltage to the gate electrode. (b) Top view of the SQPC. The width of the sample is $W=10$\,$\mu$m, the length $L=400$\,nm and the width between the two gate electrodes is $W_{g}=100$\,nm}
	\label{sample1}
\end{figure}

It is well known that the critical supercurrent $I_{C}$ of a classical Josephson point contact is directly related to its conductance $G$ and is given by $\pi G\Delta_{0}/e$,\cite{kul78} where $\Delta_{0}$ is the energy gap of the superconductor. The same holds for a SQPC.
The conductance of a QPC is given by the well known Landauer B\"{u}ttiker formula
\begin{equation}
	G=\frac{2e^{2}}{h}\sum^{N}_{n=1}T_{n}
	\label{cond}
\end{equation}
where $N=2W_{g}/\lambda_{F}$ is the maximum number of the 1D transport modes in the quantum point contact and $n\leq N$ is the index for each transport mode. The transmission coefficient for the $n$-th transport mode is given by $T_{n}$. In the case of ballistic electron transport in the quantum point contact all transport modes with $n\leq N$ are completely open $T_{n}=1$. Each open transport mode contributes the quantized conductance $\Delta G_{0}=2e^2/h$ to the total conductance 

\begin{equation}
  G=N\frac{2e^{2}}{h}=N\Delta G_{0}\;.
  \label{cond1} 
\end{equation}

Eq. \ref{cond1} tells us that the conductance changes stepwise with stephight $\Delta G_{0}$ as a function of the width of the constriction as the maximum number of transport modes $N=2W_{g}/\lambda_{F}$, which is an integer, depends on the constriction width. One has to point out that the conductance quantum $\Delta G_{0}$ does not depend on the geometry of the conductor. 

In the following an ideal interface between the superconductor and the normal conductor is assumed, i.e. there is no potential barrier at the interface. In accordance with Ref. \onlinecite{btk82} we will use $Z$ to describe the barrier strength. $Z$ is related to the transmission probability $T_{Z}$ of the interface barrier by $T_{Z}=1/(1+Z^2)$.  
In the case of a short SQPC ($\xi_{0}\gg L$) and $Z=0$ the supercurrent is given by\cite{bee91}
\begin{equation}
  I_{C}=N\frac{2e^2}{h}\frac{\pi\Delta_{0}}{e}=G\frac{\pi\Delta_{0}}{e}\;.
  \label{supershort}
\end{equation}

All N transport modes carry the quantized supercurrent $\Delta I_{C0}=e\Delta_{0}/\hbar$ 
which does not depend on the junction geomety. As for the conductance the supercurrent changes stepwise as a function of the width of the constriction.
 
In the opposite case of a long SQPC ($L\geq\xi_{0}$) the Josephson current flows via many bound states and the quantization of the supercurrent is not any more universal but depends on junction parameters. \cite{fur92} The ratio $L/\xi_{0}$ gives roughly the number of Andreev bound states \cite{and64} within the energy gap $\Delta_{0}$ which carry the Josephson current. \cite{jos62} In this case and assuming no barrier potential at the interface ($Z=0$) the supercurrent is quantized in units of $e/(\tau_{0}+\hbar/\Delta_{0})$.\cite{cht00} Here $\tau_{0}$ is the time of flight a quasiparticle requires to traverse the normal region of length $L$. During the time $\hbar/\Delta_{0}$ an electron wave packet is Andreev reflected into a hole wave packet. For a completely open transport mode the travel time can be approximated by $\tau_{0}=L/v_{F}$ and the supercurrent quantization saturates at the nonuniversal value\cite{cht00} 

\begin{equation}
\Delta I_{C0}=\frac{ev_{F}}{L+\pi\xi_{0}}\;.
\end{equation}

In contrast to the supercurrent quantization in the short junction limit ($\xi_{0}\gg L$) (Eq. \ref{supershort}) the supercurrent quantization in the long junction limit ($L\geq\xi_{0}$) depends both on the Fermi velocity $v_{F}$ and the junction length $L$ of the normal conducting region.
A finite barrier potential at the interface between the superconductor and normal conductor ($Z>0$) and a Fermi velocity mismatch will further decrease the probability of Andreev reflection and increase the probability of normal reflection.\cite{btk82} This will influence both the conductance and the supercurrent quantization. In the case of the conductance, a quasiparticle which travelled through the constriction will have a finite probability to be reflected at the normal conductor/superconductor interface and backscattered through the constriction in the opposite direction. This results in a transmission probability $T_{n}$ in Eq.\ref{cond} smaller than 1 even if the transport through the constriction itself is purly ballistic. Consequently the conductance quantization will not have the universal value $2e^2/h$ and the conductance as a function of the constriction width will have stephights depending on the random transmission probabilities $T_{n}$.  
The effect of reduced Andreev probability will be a reduction of the supercurrent through the constriction and consequently a reduction of the quantization of the supercurrent.\cite{fur92,shc00,cht00}

\section{Sample}

The schematic cross section and top view of the sample is shown in Fig. \ref{sample1}.
The 2DEG is localized in a 4\,nm thick InAs layer inserted in an In$_{0.52}$Al$_{0.48}$As/In$_{0.53}$Ga$_{0.47}$As heterostructure grown by molecular beam epitaxy on a Fe doped semi-insulating InP substrate. The two 100 nm thick Nb electrodes, which are coupled to the 2DEG, were defined by lift off process and electron beam lithography. InAs was used as it does not form a Schottky barrier at the interface to the niobium electrodes compared to GaAs. 
Details of the fabrication process are reported elsewhere. \cite{tak95} The distance between the Nb electrodes is $L=400$\,nm and the width of the junction is $W=10$\,$\mu$m. 
Shubnikov-de Haas measurements\cite{tak95b} of the 2DEG at 4.2 K on similar samples give the sheet carrier concentration $n_{S}=2.3\times 10^{12}$\,cm$^{-2}$, the mobility $\mu =111000$\,cm$^{2}$/Vs and the effective mass $m^{\ast}=0.045m_{e}$, where $m_{e}$ is the free electron mass. This relsults in a fermi velocity $v_{F}=\sqrt{2\pi\hbar^{2}n_{S}/{m^{\ast}}^{2}}=9.8\times 10^{5}$\,m/s and elastic scattering time $\tau=\mu m^{\ast}/e=2.84\times 10^{-12}$\,s, where e is the elementary electric charge. From these values the mean free path $l=v_{F}\tau=2.8$\,$\mu$m and the normal coherence length in the clean limit $\xi_{N}=\hbar v_{F}/2\pi k_{B}T=0.28$\,$\mu$m at 4.2\,K are calculated. The Fermi wave length is $\lambda_{F}=\sqrt{2\pi/n_{S}}=16.5$\,nm. 

The length of the Al split gate is 100\,nm and the distance between the two gate electrodes is $W_{g}=100$\,nm. By applying a gate voltage $V_{g}\simeq -1$\,V the 2DEG under the gate electrodes is depleted. In this case the current is flowing only within the constriction between the two gate electrodes. Going to more negative gate voltages $V_{g}<-1$\,V will further deplete the 2DEG within the constriction reducing the width of the constriction $W_{g}$. Eventually at gate voltages $V_{g}<-2.1$\,V the 2DEG within the constriction is also completely pinched off. 

The Nb electrodes have a superconducting transition temperature of about 6\,K which results in an energy gap $\Delta_{0}\simeq 0.9$\,meV. This gives us for the superconducting coherence length in the 2DEG $\xi_{0}=\hbar v_{F}/\pi\Delta_{0}=230$\,nm where $v_{F}$ is the Fermi velocity in the 2DEG. Therefore the junction is in the ballistic ($l>L$) and long junction ($L\geq\xi_{0}$) regime and in the clean limit ($\xi_{N}<l$).

At a temperature of 25\,mK and at zero gate voltage the junction under investigation has a critical current $I_{C0}=8.5$\,$\mu$A and a normal state resistance $R_{N}=38$\,$\Omega$.

\section{Measurements}
The measurements were performed in a $^{3}$He-$^{4}$He dilution refrigerator with base temperature of 15\,mK. To protect the sample from external noise and from 4\,K photons the electrical lines to the sample in the cryostat are well filtered. At the 1K-pot a home-built RCL filter \cite{bla03} with a cut-off frequency of 100\,MHz is installed. At the mixing chamber a combination of two meters Thermocoax \cite{zor95} plus a home-built copper-powder filter is installed with a cut off frequency of 1\,GHz.

\begin{figure}[htb]
	\begin{center}
		\includegraphics[width=8cm]{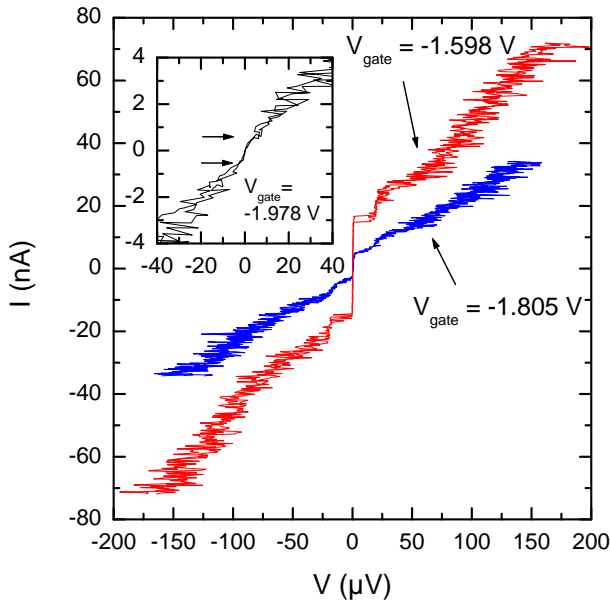}
	\end{center}
	\caption{Measured current voltage characteristics (CVC) at $T=25$\,mK for different gate voltages. In the inset the arrows indicate the onset of supercurrent at $V_{gate}=-1.978$\,V which is carried by the first open transport mode.}
	\label{IcRnIV}
\end{figure}

The current voltage characteristics (CVC) of the SQPC were recorded using standard four point measurement technique. The CVC for three different gate voltages at $T=25$\,mK are shown in Fig. \ref{IcRnIV}. For increasing absolute value of the gate voltage $\left| V_{g}\right|$ the supercurrent is decreasing. For gate voltages $V_{g}<-1.8$\,V the supercurrent branch in the current voltage characterisics shows a finite resistance. The critical currents at those gate voltages are in the range of $1$\,nA which correspond to a Josephson energy $E_{J}/k_{B}=I_{C}\Phi_{0}/2\pi k_{B}=24$\,mK, where $\Phi_{0}=h/2e$ is the superconducting flux quantum and $k_{B}$ is the Boltzmann constant. Therefore the finite resistance can be explained by thermal activation of the phase diffusion accross the junction which results in a finite voltage accross the junction in the supercurrent branch of the current voltage characteristic.\cite{amb69}

\begin{figure}[htb]
	\begin{center}
		\includegraphics[width=8cm]{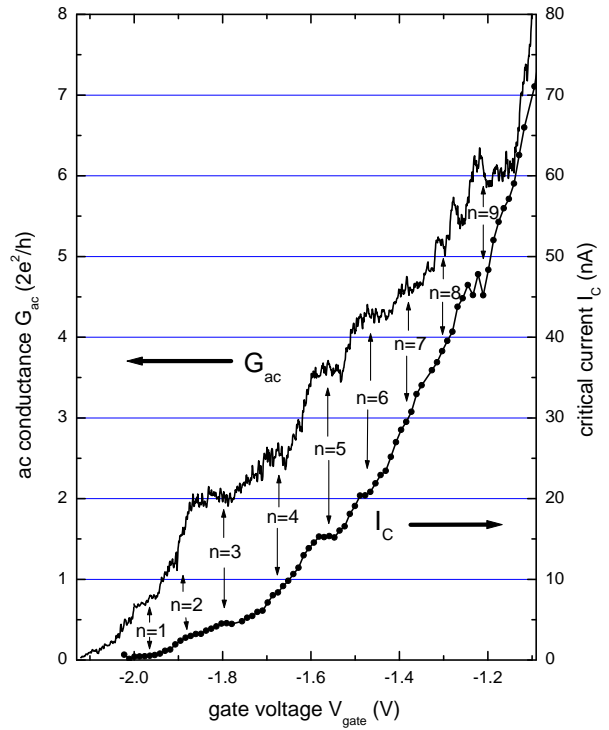}
	\end{center}
	\caption{Critical current $I_{C}$ and differential zero bias conductance $G_{ac}$ as a function of the gate voltage $V_{g}$ at $T=25$\,mK. The mode number is indicated by $n$.}
	\label{IcRnsteps}
\end{figure}	

The critical current values which where extracted from the CVC for gate voltages $-2.1$\,V$<V_{g}<-1.1$\,V are plotted in Fig. \ref{IcRnsteps} together with the differential zero bias conductance $G_{ac}$ at a magnetic field of $200$\,mT to suppress the Josephson supercurrent. The differential conductance was measured with an AC excitation current $I_{AC}=10$\,nA. To check for nonlinearities possibly due to a residual Josephson current in the current voltage charachterisitics at a magnetic field of $200$\,mT we varied the AC excitation current from 5\,nA to 50\,nA which results in a voltage drop accross the junction from 90\,$\mu$V to 900\,$\mu$V at $V_{g}=-1.98$\,V and from 11\,$\mu$V to 110\,$\mu$V at $V_{g}=-1.2$\,V. No difference within the measurement accuracy between the differential conductances as a function of the gate voltage for the different excitation currents was observed. Herewith we can rule out any influence of the Josephson current on the measured zero bias conductance. 

In Fig.\ref{IcRnsteps} one can clearly see that both the conductance $G_{ac}$ and the supercurrent $I_{C}$ of the SQPC change stepwise as a function of the gate voltage $V_{g}$. The steps are marked with the index $n$ which corresponds also to the transport mode index contributing both to the conductance and the supercurrent. The pronounced steps ($n=1,3,5,6,9$) in the conductance are also seen as pronounced steps in the supercurrent. This shows the direct correlation between the supercurrent and the conductance. In particular one can see that as soon as the first ($n=1$) transport mode contributes to the first conductance step at a gate voltage $V_{g}=-1.98$\,V it also contributes to the first supercurrent step. The current voltage characteristic is shown in the inset of Fig.\ref{IcRnIV}. This is the first time that a supercurrent which is carried by a single transport mode has been observed experimentally in a SQPC. 

\section{discussion}

The average hight of the steps of the conductance as a function of the gate voltage in Fig.\ref{IcRnsteps} is approximately $\Delta G_{eff}=0.66\Delta G_{0}$, which gives us an effective average transmission probability
$T_{eff}=\Delta G_{eff}/\Delta G_{0}=0.66$ for a quasiparticle travelling through the constriction. This can be attributed to the finite barrier potential at the 2DEG-Nb interface. A rough estimate\cite{oct83} of the the barrier strength $Z$ at the interface can be derived from the relation $R_{N}=R_{Sh}(1+2Z^2)$ where $R_{N}$ is the resistance of the SQPC at zero gate voltage and $R_{Sh}$ the Sharvin resistance $R_{Sh}=(h/2e^2)(\lambda_{F}/2W)$. With $W=10\,\mu$m and $\lambda_{F}=16.5$\,nm we get for the Sharvin resistance $R_{Sh}=10.7\,\Omega$ which results in a barrier strength $Z=1.1$ or transmission probability $T_{Z}=0.45$.
The reason for $T_{Z}$ being smaller than $T_{eff}$ can be attributed to the effect, that not all quasiparticles which are normal reflected from the 2DEG/Nb interface will be scattered back through the constriction. There could also be an inhomogeneity along the interface which will result in a varying transmission probability along the interface. $T_{Z}$ is an average transmission probability along the junction width $W=10$\,$\mu$m and $T_{eff}$ is a local measure of the transmission probability on the length scale of the constriction width $W_{g}=100$\,nm which can differ substantially from $T_{Z}$. Another reason could be that the zero bias conductance at $200$\,mT is still affected by Andreev reflection at the 2DEG/Nb interface.\cite{tak95b} 
 
The hight of the steps of the supercurrent $I_{C}$ as a function of gate voltage $V_{g}$ in Fig.\ref{IcRnsteps} depends on the gate voltage with increasing step hight for increasing step number. The step hight for $n\leq 3$ is smaller than $2$\,nA, which could be attributed to a reduced measured critical current due to thermal fluctuations and spurious noise in the measurement setup. From step $n=3$ to $n=6$ we have a step hight of $\Delta I_{c}\simeq 5$\,nA. From step $n=6$ to step $n=9$ the current increases by $25$\,nA, which would result in a step hight $\Delta I_{C}\simeq 8.5$\,nA, but unfortunately no clear steps for $n=7$ and $n=8$ are seen.

According to Ref.\cite{shc00} the supercurrent step hight in the presence of scattering at the 2DEG-Nb interface is given by $\Delta I_{C0}=T_{Z}ev_{F}/4\pi(L+\pi\xi_{0})=5$\,nA in the long junction limit ($L>\xi_{0}$) and $\Delta I_{C0}=T_{Z}^{2}e\Delta_{0}/8\hbar=5.5$\,nA in the short junction limit ($L<\xi_{0}$). These values do not take any resonances in the supercurrent due to normal reflection at the 2DEG-Nb interface into account. This seems to be the case in our sample as we do not see any resonances in the supercurrent as a function of the gate voltage. 
From the critical current at zero gate voltage $I_{C0}=8.5\mu$A which is carried by $N=2W/\lambda_{F}\simeq 1200$ transport modes we would expect an average critical current step hight $\Delta I_{C}=I_{C0}\lambda_{F}/2W\simeq 7$\,nA, which agrees with the order of magnitude of the supercurrent step hights we have measured.

\section{conclusion}

We have measured the conductance and supercurrent quantization of a S-2DEG-S Josephson junction SQPC. By applying a gate voltage to the split gate the current was forced to flow within a constriction of width $W<100$\,nm in the 2DEG. Both the conductance and the supercurrent as a function of gate voltage showed a pronounced steplike structure. For the first time it could be shown that the supercurrent and the conductance are directly correlated. Especially we could detect a supercurrent in the SQPC which was carried by the first open transport mode in the SQPC.

\begin{acknowledgments}
We would like to acknowledge stimulating discussions with N. Chtchelkatchev, V. Shumeiko and F. Lombardi.
This work was supported by the Japanese NEDO, G. Gustavsson foundation and Deutsche Forschungsgemeinschaft.
\end{acknowledgments}


\end{document}